\documentclass[preprint,showpacs,preprintnumbers,amsmath,amssymb]{revtex4}
\usepackage{graphicx}
\usepackage{dcolumn}
\usepackage{bm}

\begin{document}
\preprint{PRC/$^{13}$N($d,n$)$^{14}$O manuscript}

\title{The $^{13}N(d,n)^{14}O$ Reaction and
the Astrophysical $^{13}N(p,\gamma)^{14}O$ Reaction Rate}
\author{Z. H. Li}\email{zhli@ciae.ac.cn}
\author{B. Guo}
\author{S. Q. Yan}
\author{G. Lian}
\author{X. X. Bai}
\author{Y. B. Wang}
\author{S. Zeng}
\author{J. Su}
\author{B. X. Wang}
\author{W. P. Liu}
\author{N. C. Shu}
\author{Y. S. Chen}
\author{H. W. Chang}
\author{L. Y. Jiang}
\affiliation{%
China Institute of Atomic Energy, P. O. Box 275(46),  Beijing 102
413, P. R. China}%


\begin{abstract}
$^{13}$N($p,\gamma$)$^{14}$O is one of the key reactions in the hot
CNO cycle which occurs at stellar temperatures around $T_9$ $\geq$
0.1. Up to now, some uncertainties still exist for the direct
capture component in this reaction, thus an independent measurement
is of importance. In present work, the angular distribution of the
$^{13}$N($d,n$)$^{14}$O reaction at $E_{\rm{c.m.}}$ = 8.9 MeV has
been measured in inverse kinematics, for the first time. Based on
the distorted wave Born approximation (DWBA) analysis, the nuclear
asymptotic normalization coefficient (ANC), $C^{^{14}O}_{1,1/2}$,
for the ground state of $^{14}$O $\rightarrow$ $^{13}$N + $p$ is
derived to be $5.42 \pm 0.48$ fm$^{-1/2}$. The
$^{13}$N($p,\gamma$)$^{14}$O reaction is analyzed with the R-matrix
approach, its astrophysical S-factors and reaction rates at energies
of astrophysical relevance are then determined with the ANC. The
implications of the present reaction rates on the evolution of novae
are then discussed with the reaction network calculations.
\end{abstract}

\maketitle
\section{Introduction}
In stellar-evolution models, hydrogen burning in massive stars
proceeds largely through the CNO cycle. For the normal CNO cycle,
the dominant sequence of reactions is
\begin{equation*}
\label{eq1}
^{12}C(p,\gamma)^{13}N(\beta^{+})^{13}C(p,\gamma)^{14}N(p,\gamma)^{15}O
(\beta^{+})^{15}N(p,\alpha)^{12}C.
\end{equation*}
When temperature increases, the $\beta^+$ decay of $^{13}$N limits
the cycle, and most of the C, N and O nuclei would be processed into
$^{13}$N. Consequently, the $^{13}$N($p,\gamma$)$^{14}$O reaction
provides a second channel for destruction of $^{13}$N, and the
dominant sequence becomes
\begin{equation*}
\label{eq2}
^{12}C(p,\gamma)^{13}N(p,\gamma)^{14}O(\beta^{+})^{14}N(p,\gamma)^{15}O
(\beta^{+})^{15}N(p,\alpha)^{12}C.
\end{equation*}
This reaction sequence is called hot or ``$\beta$-limited'' CNO
cycle, and the $\beta^+$ decays of $^{14}$O and $^{15}$O limit this
cycle. The CNO cycles convert four hydrogen nuclei into an alpha
particle and the energy release in the cycles is about 26.7 MeV,
which is the important source of stellar energy
generation~\cite{Mat84}. Since the $\beta^+$ decays of $^{14}$O and
$^{15}$O are much quicker than that of $^{13}$N, the hot CNO cycle
should produce energy much faster than the normal CNO cycle. Hence,
a rapid change of the temperature dependent energy generation rate
occurs when the CNO cycle transits from the normal one to the hot
one. $^{13}$N($p,\gamma$)$^{14}$O is one of the important reactions
which controls this transition~\cite{Wie99}. Therefore precise
determination of the rates for the $^{13}$N proton capture reaction
is vital for determining the transition temperature and density
between the normal and hot CNO cycles.

At the energies of astrophysical interest, the
$^{13}$N($p,\gamma$)$^{14}$O reaction is dominated by the low energy
tail of the $s$-wave capture on the broad 1$^-$ resonance at $E_r$ =
527.9 keV (which has a total width of 37.3 $\pm$ 0.9 keV). A
considerable effort has been expended in recent years to determine
the parameters for the resonance. These include the direct
measurements using the radioactive $^{13}$N beam \cite{Dec91,Del93},
particle transfer reactions \cite{Chu85,Fer89,Smi93,Mag94}, and
Coulomb dissociation of high-energy $^{14}$O beams in the field of a
heavy nucleus \cite{Bau94,Mot91,Kie93}. The direct capture
contribution is significantly smaller than the contibution due to
the tail of the resonance within the Gamow window. But since both
resonant and non-resonant captures proceed via $s$-waves and then
decay by E1 transitions, there is an interference between the two
components. Thus the capture reaction within the Gamow window can be
enhanced through constructive interference or reduced through
destructive interference. The non-resonant component of the cross
section has been calculated by several groups, either separately or
as part of the calculation of the total cross
section~\cite{Mat84,Bar85,Fun87,Des89}. Since there are significant
differences among the various calculations, the determination of the
 $^{13}$N($p,\gamma$)$^{14}$O direct capture component through an
independent approach is greatly needed. A practicable method is to
extract the direct capture cross section of the
$^{13}$N($p,\gamma$)$^{14}$O reaction using the direct capture model
\cite{Rol73,Chr61} and the spectroscopic factor (or ANC), which can
be deduced from the angular distribution of one proton transfer
reaction. Decrock et al. extracted the spectroscopic factor for
$^{14}$O $\rightarrow$ $^{13}$N + $p$ from the
$^{13}$N($d,n$)$^{14}$O cross section \cite{Dec93}. Tang et al.
derived the ANC for $^{14}$O $\rightarrow$ $^{13}$N + $p$ from the
$^{14}$N($^{13}$N,$^{14}$O)$^{13}$C angular distribution
\cite{Tang04}. The S-factors for the direct capture of the
$^{13}$N($p,\gamma$)$^{14}$O reaction from these two works differ
from each other by a factor of 30{\%}. Thus, further measurement is
important for the determination of the spectroscopic factor (or ANC)
for $^{14}$O $\rightarrow$ $^{13}$N + $p$ and the astrophysical
S-factor of the $^{13}$N($p,\gamma$)$^{14}$O reaction.

In the present work, we have measured the angular distribution of
the $^{13}$N($d,n$)$^{14}$O reaction at $E_{\rm{c.m.}}$ = 8.9 MeV in
inverse kinematics. The spectroscopic factor and ANC were derived
based on distorted wave Born approximation (DWBA) analysis, and used
to calculate the astrophysical S-factors and rates of
$^{13}$N($p,\gamma$)$^{14}$O direct capture reaction at energies of
astrophysical interest with the R-matrix approach. We have also
computed the contribution from the resonant capture and the
interference effect between resonant and direct capture. The total
reaction rates are then used in the reaction network calculations at
the typical density and temperature of novae environment.

\section{measurement of the $^{13}$N($d,n$)$^{14}$O angular distribution}

The experiment was carried out using the secondary beam facility
\cite{Bai95, Liu03} of the HI-13 tandem accelerator, Beijing. An 84
MeV $^{12}$C primary beam from the tandem impinged on a 4.8 cm long
deuterium gas cell at a pressure of 1.6 atm. The front and rear
windows of the gas cell are Havar foils, each in thickness of 1.9
mg/cm$^2$. The $^{13}$N ions were produced via the $^{2}$H($^{12}$C,
$^{13}$N)$n$ reaction. After the magnetic separation and focus with
a dipole and a quadruple doublet, the secondary beam was further
purified with a wien filter. The 69 MeV $^{13}$N secondary beam was
then delivered with typical purity of 92\%. The main contaminants
were $^{12}$C ions out of Rutherford scattering of the primary beam
in the gas cell windows as well as on the beam tube. The $^{13}$N
beam was collimated with two apertures in diameter of 3 mm and
directed onto a (CD$_{2})_{n}$ target in thickness of 1.5 mg/cm$^2$
to study the $^{2}$H($^{13}$N,$^{14}$O)$n$ reaction. The typical
beam intensity and beam energy spread on the target were 1500 pps
and 1.8 MeV FWHM for long-term
 measurement, respectively.  A carbon target in thickness of $1.5$ mg/cm$^2$ served as the background measurement.

\begin{figure}
\includegraphics[height=8 cm,angle=-90]{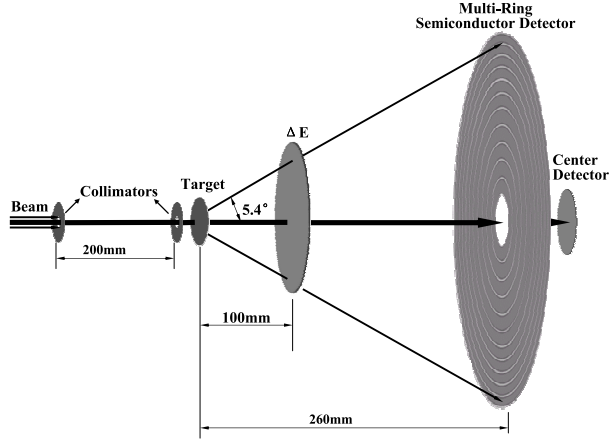}
\caption{\label{fig:setup}Schematic layout of the experimental
setup}
\end{figure}

The experimental setup is shown in Fig.~\ref{fig:setup}. A 300
$\mu$m thick Multi-Ring Semiconductor Detector (MRSD) with center
hole was used as a residue energy ($E_\mathrm{r}$) detector which
composed a $\Delta E - E_{\mathrm{r}}$ counter telescope together
with a 21.6 $\mu$m thick silicon $\Delta E$ detector and a 300
$\mu$m thick silicon center $E_{r}$ detector. Such a detector
configuration covered the laboratory angular range from $0^\circ$ to
$5.4^{\circ}$, and the corresponding angular range in the
center-of-mass frame is from $0^\circ$ to $66.5^{\circ}$. This setup
also facilitates to precisely determine the accumulated quantity of
incident $^{13}$N because the $^{13}$N themselves are recorded by
the counter telescope simultaneously.

\begin{figure}
\includegraphics[height=6 cm]{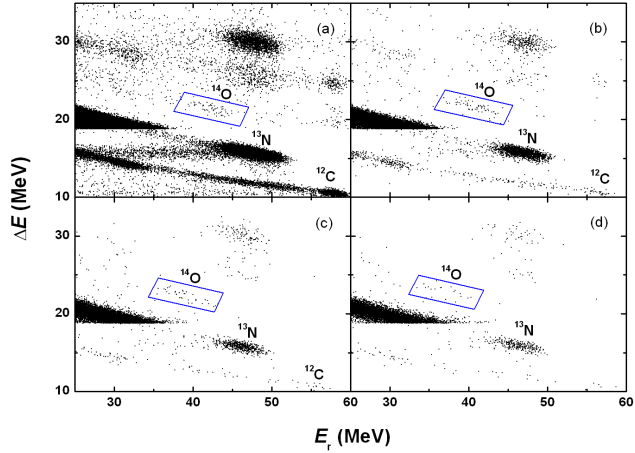}
\caption{\label{fig:deer}(Color online) Scatter plots of  $\Delta E$
$vs.$ $E_{r}$ for the $^{13}N(d,n)^{14}O$ reaction measurement with
$(CD_{2})_{n}$ target. (a) - (d) display the $\Delta E$- $E_{r}$
spectra for the first four rings of MRSD.}
\end{figure}

\begin{figure}
\includegraphics[height=6 cm]{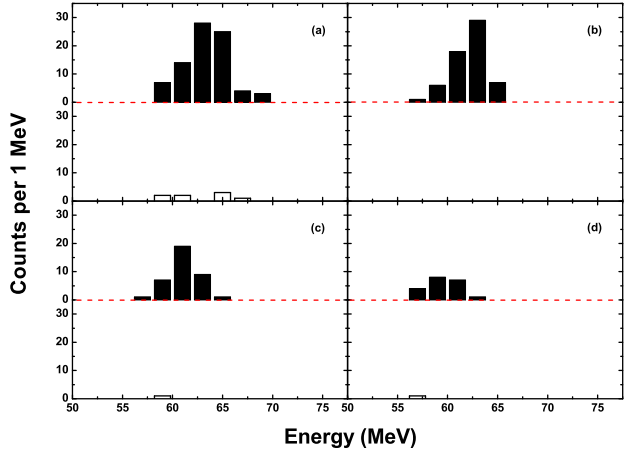}
\caption{\label{fig:effbg}(Color online) Comparison of total energy
spectra between (CD$_2$)$_n$ and pure Carbon target. (a)-(d)
represent the spectra for ring 1,2,3 and 4. The solid bar and the
empty bar stand for the $^{14}$O spectra from (CD$_2$)$_n$ target
and carbon target, respectively.}
\end{figure}

The accumulated quantity of incident $^{13}$N is approximately
$3.54\times 10^8$ for the $(CD_{2})_{n}$ target measurement, and
1.18 $\times$ $10^{8}$ for background measurement with the carbon
target. Fig.~\ref{fig:deer} (a) - (d) display the $\Delta E-E_{r}$
scatter plots for the first four rings, respectively. For the sake
of saving CPU time in dealing with the experimental data, we set a
cut line of $\Delta E$ = 19 MeV. All the events below the line are
scaled down by a factor of 1000, and the $^{14}$O events are not
affected by this cut. The four two-dimension gates plotted in Fig.
~\ref{fig:deer} (a) - (d) are the $^{14}$O kinematics regions based
on the Monte Carlo simulation, taking the beam spot size, energy
spread, angular divergence and the target thickness into account.
The $^{14}$O events can be clearly identified through this figure.
Fig. ~\ref{fig:effbg} displays the comparison of the events from
(CD$_2$)$_n$ target with the background from carbon target in the
$^{14}$O kinematics regions for the first four rings. The background
events in the first ring of MRSD mainly come from the pileup of
$^{12}$C contaminants in the beam, they disappear in the outer
rings. After the background subtraction, the angular distribution in
center of mass frame for the forward angles is given in
Fig.~\ref{fig:andis}. The uncertainties of differential cross
section mainly arise from the statistics, the assignment of $^{14}$O
kinematics regions, the uncertainties of the target thickness and
the solid angle. The angular uncertainties include the random
reaction point in the target, the angular uncertainty of the
secondary beam, the angular straggling of $^{13}$N and $^{14}$O in
the target and the $\Delta E$ detector. The total angular error for
each ring is about 0.6 degree, less than the width of each ring.

\section{Determination of the $^{14}$O Nuclear ANC}

The spins and parities of $^{13}$N and $^{14}$O (ground state) are
$1/2^{-}$ and $0^{+}$, respectively. The cross section of the
$^{13}$N($d, n$)$^{14}$O reaction is dominated by the $s$-wave
proton transition to $1p1/2$ orbit in $^{14}$O ground state. If the
reaction is peripheral, the differential cross section can be
expressed as
\begin{equation}\label{eq4}
({d\sigma \over d\Omega})_{exp} = (\frac{C_d}{b_d})^2
(\frac{C^{^{14}O}_{1,1/2}}{b^{^{14}O}_{1,1/2}})^2
\sigma_{1,1/2}(\theta),
\end{equation}
where $({d\sigma \over d\Omega})_{exp}$ and $\sigma_{l,j}(\theta)$
denote the measured and theoretical differential cross sections
respectively. $C^{^{14}O}_{1,1/2}$ and $C_d$ stand for the nuclear
ANCs for the $^{14}$O $\rightarrow$ $^{13}$N + $p$ and $d$
$\rightarrow$ $n$ + $p$ virtual decays, $b^{^{14}O}_{1,1/2}$ and
$b_d$ being the single particle ANCs of the bound state protons in
$^{14}$O and deuteron. By knowing the value of $C_d$ , the
$C^{^{14}O}_{1,1/2}$ can then be extracted by normalizing the
theoretical differential cross sections to the experimental data by
Eq. (\ref{eq4}).
\begin{figure}
\includegraphics[height=6 cm]{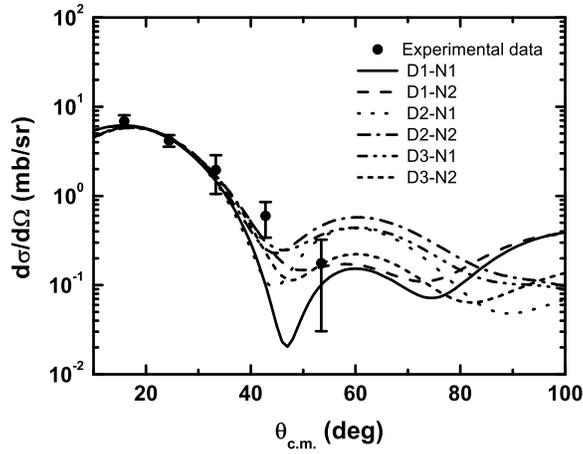}
\caption{\label{fig:andis}The angular distribution of the
$^{13}N(d,n)^{14}O$ reaction at $E_{c.m.}$ = 8.9 MeV, together with
DWBA calculations using different optical potential parameters.}
\end{figure}

\begin{figure}
\includegraphics[height=6 cm]{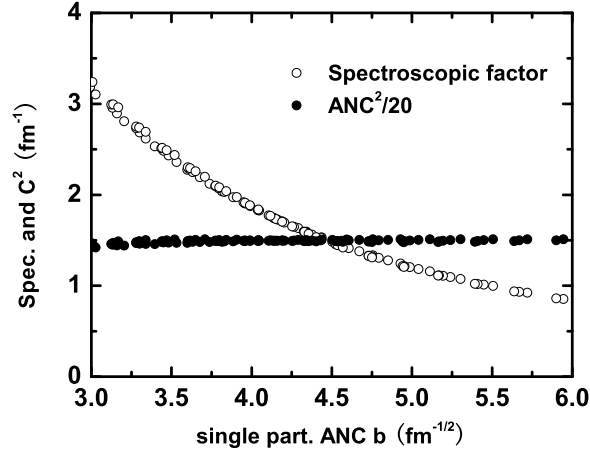}
\caption{\label{fig:sanc}Comparison of spectroscopic factors with
the ANCs derived from the present experiment for different
geometries of the Woods-Saxon potentials.}
\end{figure}

The DWBA code DWUCK \cite{Kun78} is used to compute the angular
distribution.  All the optical potential parameters for the entrance
channel are taken from Ref. \cite{Per76}, the ones for the exit
channel are from Refs. \cite{Per76} and \cite{Wat69}, respectively,
these parameters are listed in Table~\ref{tab:table1}. In the
present DWBA calculation, the differential cross sections at three
forward angles are used to extract the ANC, and $C_{d}$ is taken to
be 0.872 fm$^{-1/2}$ from Ref. \cite{Blo77}. The normalized angular
distributions from the six sets of optical potential parameters are
also presented in Fig.~\ref{fig:andis}, each curve corresponds to
one nuclear ANC, $C^{^{14}O}_{1,3/2}$, the spectroscopic factor is
calculated with $C^2/b^2$. The nuclear ANCs and the spectroscopic
factors deduced from the present experimental data are listed in
Table \ref{tab:table2}, the average values of them are 5.42 $\pm$
0.48 fm$^{-1/2}$ and 1.88 $\pm$ 0.34, respectively. The present ANC
accords with the result extracted from the $^{14}$N($^{13}$N,
$^{14}$O)$^{13}C$ transfer reaction by Tang et al. \cite{Tang04},
and the present spectroscopic factor is larger than the previous one
(0.9) extracted from the total cross section of $^{13}$N($d,
n$)$^{14}O$ at lower energy \cite{Dec93}. The uncertainties of the
nuclear ANC and the spectroscopic factor are mainly from the
difference of the calculated angular distributions with different
optical potentials, as well as the experimental errors. Since we do
not measure the optical potential parameters and used six sets of
potential parameters from the neighboring nuclei, the error bar of
present work is a bit larger than that of Ref.~\cite{Tang04}.
Fig.~\ref{fig:sanc} shows the comparison of the spectroscopic
factors with ANCs of $^{14}$O $\rightarrow$ $^{13}$N + $p$ from the
different geometry parameters of the Woods-Saxon potential for the
single particle bound state (by changing the radius and diffuseness
$r_0$ and $a$). One can see that the spectroscopic factors vary
strikingly, while the ANCs are nearly a constant, thus indicating
that the $^{13}$N($d, n$)$^{14}$O reaction at present energy is
dominated by peripheral process.

\begin{table}
\caption{\label{tab:table1} Optical potential parameters used in
DWBA calculations, where $V$ , $W$ are in MeV, $r$ and $a$ are in
fm, the geometrical parameters of single particle bound state are
set to be $r_0$ = 1.25 fm and $a$ = 0.65 fm. D1, D2 and D3
correspond to the optical potentials for $d$ + $^{13}$N, and N1, N2
represent the ones for $n$ + $^{14}$O.}
\begin{ruledtabular}
\begin{tabular}{ccccccc}
Channel &\multicolumn{3}{c}{Entrance}& &\multicolumn{2}{c}{Exit}\\
 & D1 & D2 & D3 & & N1 & N2 \\
\hline
$V_r$    & 117.9& 116.0 & 130.4 & & 49.2 & 61.56\\
$r_{0r}$ & 0.81 &  1.0  &  0.9  & & 1.2  &  1.14\\
$a_r$  & 1.07  & 0.8    & 0.9   & & 0.65 & 0.57 \\
$W_V$  &       & 4.13   &      & &      &       \\
$r_{wv}$&      & 1.0    &      & &      &       \\
$a_{wv}$&      & 0.8    &      & &      &       \\
$W_s$ &19.61   & 4.13   & 6.63 & & 6.0  &  7.74 \\
$r_{0s}$& 1.84 & 2.0    & 1.90 & & 1.2  & 1.14  \\
$a_s$ & 0.35   & 0.6    & 0.56 & & 0.47 &  0.5  \\
$V_{SO}$ &     & 6.76   &      & & 7.0  &  5.5  \\
$r_{0SO}$ &    & 1.0    &      & & 1.20 & 1.14  \\
$a_{SO}$ &     & 0.8    &      & & 0.65 & 0.8   \\
$r_{0c}$ & 0.81& 1.5    & 1.30 & &      &       \\
\end{tabular}
\end{ruledtabular}
\end{table}

\begin{table}
\caption{\label{tab:table2} The $^{14}$O nuclear ANC,
$C^{^{14}O}_{1,1/2}$, and spectroscopic factor,
$S_{1,1/2}^{^{14}O}$, deduced from the angular distribution of the
$^{13}$N($d, n$)$^{14}O$ reaction using the combination of optical
potentials for the entrance and exit channels.}
\begin{ruledtabular}
\begin{tabular}{ccc}
optical & {$C_{1,1/2}^{^{14}O}$}& $S_{1,1/2}^{^{14}O}$\\
potentials &  (fm$^{-1/2}$)  &    \\
\hline
D1-N1     &  5.27 $\pm$ 0.42   &  1.77 $\pm$ 0.28  \\
D1-N2     &  4.95 $\pm$ 0.17   &  1.56 $\pm$ 0.11  \\
D2-N1     &  6.02 $\pm$ 0.61   &  2.31 $\pm$ 0.47  \\
D2-N2     &  5.42 $\pm$ 0.29   &  1.87 $\pm$ 0.20  \\
D3-N1     &  5.56 $\pm$ 0.27   &  1.97 $\pm$ 0.19  \\
D3-N2     &  5.31 $\pm$ 0.19   &  1.80 $\pm$ 0.13  \\
\hline
average &  5.42 $\pm$ 0.48 & 1.88 $\pm$ 0.34 \\
\end{tabular}
\end{ruledtabular}
\end{table}

\section{Astrophysical S-factor of $^{13}$N($p,\gamma$)$^{14}$O}

According to the traditional direct capture model
\cite{Rol73,Chr61,Lizh05}, the direct capture of the
$^{13}$N($p,\gamma$)$^{14}$O reaction is believed to be dominated by
the $E1$ transition from incoming $s$ wave to bound $p$ state. The
direct capture cross section can be computed by
\begin{equation}
\label{eq5}
\sigma_{dc}=\frac{16\pi}{9}k_{\gamma}^3\bar{e}^2A_{ij}S_{l,j}
|\int^\infty_0drr^2\varphi_{l_f}(r)\psi_{l_i}(r)|^2 .
\end{equation}
where $k_{\gamma}=\epsilon_{\gamma}/\hbar c$ is the wave number of
the emitted $\gamma$-ray (of energy $\epsilon_{\gamma}$) and
$\bar{e} = eN/A$ is the E1 effective charge for protons, $A_{ij}$
corresponds to the angular part depending on the initial and final
angular momenta of the transition, $S_{l,j}$ is the spectroscopic
factor of the configuration $^{14}$O $\rightarrow$ $^{13}$N + $p$,
$\varphi_{l_f}(r)$ is the bound state wave function of the relative
motion of $p+^{13}$N in $^{14}$O calculated in the Woods-Saxon
potential, $\psi_{l_i}(r)$ is the optical model scattering wave
function of the colliding proton and $^{13}$N. If the spectroscopic
factor $S_{l,j}$ is deduced from the $^{13}$N($d, n$)$^{14}$O
transfer reaction, the $^{13}$N($p,\gamma$)$^{14}$O cross section
can then be calculated by Eq.~(\ref{eq5}).

\begin{figure}
\includegraphics[height=6 cm]{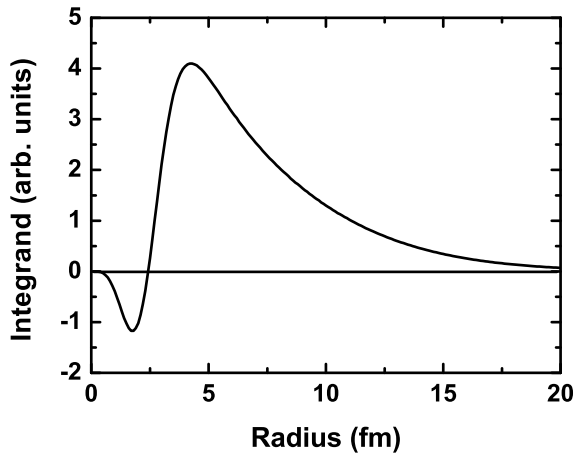}
\caption{\label{fig:integrand}The integrand of the $E$1 transition
matrix element based on a single-particle model at resonant energy.}
\end{figure}

However, this is not the case here, as a result of the tight binding
of the last proton in $^{14}$O, the contribution to the
$^{13}$N($p,\gamma$)$^{14}$O direct capture reaction at small $r$ in
Eq.~(\ref{eq5}) is important. The integrand of the $E$1 transition
matrix element at resonant energy is calculated based on a
single-particle model, as shown in Fig.~\ref{fig:integrand}. One can
see that the contribution at small $r$ is of significance,  the
simple direct capture model may be not valid due to the many
particle effects. In this case, the integral is very sensitive to
the optical potential parameters and the spectroscopic factor
required for Eq.~(\ref{eq5}) has significant uncertainties, as can
be seen from Fig.~\ref{fig:sanc}.

In this work, we will use R-matrix method to avoid the above
problems. For the radiative capture reaction $B + b \rightarrow A +
\gamma$, the R-matrix radiative capture cross section to a state of
nucleus A with a given spin $J_f$ may be written as \cite{Bar91}
\begin{equation}
\label{eq6} \sigma_{J_f}=\sum_{J_i} \sigma_{J_iJ_f},
\end{equation}

\begin{equation}
\label{eq7}
\sigma_{J_iJ_f}=\frac{\pi}{k^2}\frac{2J_i+1}{(2J_b+1)(2J_B+1)}\sum_{Il_i}|U_{Il_iJ_fJ_i}|^2.
\end{equation}
Here $J_i$ is the total angular momentum of the colliding nuclei B
and b in the initial state, $J_b$ and $J_B$ are the spins of nuclei
b and B, and $I$, $k$, and $l_i$ are their channel spin, wave number
and orbital angular momentum in the initial state. $U_{Il_iJ_fJ_i}$
is the transition amplitude from the initial continuum state
($J_i,I,l_i$) to the final bound state ($J_f,I$). In the one-level,
one-channel approximation, the resonant amplitude for the capture
into the resonance with energy $E_{R_n}$ and spin $J_i$, and
subsequent decay into the bound state with the spin $J_f$ can be
expressed as
\begin{equation}
\label{eq8}
U^R_{Il_iJ_fJ_i}=-ie^{i(\omega_{l_i}-\phi_{l_i})}\frac{[\Gamma^{J_i}_{bIl_i}(E)\Gamma^{J_i}_{\gamma
J_f}(E)]^{1/2}}{E - E_{R_n} + i \frac{\Gamma_{J_i}}{2}}.
\end{equation}
Here we assume that the boundary parameter is equal to the shift
function at resonance energy and $\phi_{l_i}$ is the hard-sphere
phase shift in the $l_i$th partial wave,
\begin{equation}
\phi_{l_i}=\arctan\Big{[}\frac{F_{l_i}(k,r_c)}{G_{l_i}(k,r_c)}\Big{]},
\end{equation}
where $F^2_{l_i}$ and $G^2_{l_i}$ are the regular and irregular
 Coulomb functions, $r_c$ is the channel radius. The Coulomb phase factor $\omega_{l_i}$ is given by
\begin{equation}
\label{eq9} \omega_{l_i}=\sum_{n=1}^{l_i}\arctan(\frac{\eta_i}{n}),
\end{equation}
where $\eta_i$ is the Sommerfeld parameter.
$\Gamma^{J_i}_{bIl_i}(E)$ is the observable partial width of the
resonance in the channel $B$ + $b$, $\Gamma^{J_i}_{\gamma J_f}(E)$
is the observable radiative width for the decay of the given
resonance into the bound state with the spin $J_f$, and
$\Gamma_{J_i} \approx \sum\limits_I \Gamma^{J_i}_{bIl_i}$ is the
observable total width of the resonance level. The energy dependence
of the partial widths is determined by
\begin{equation}
\label{eq10}
\Gamma^{J_i}_{bIl_i}(E)=\frac{P_{l_i}(E)}{P_{l_i}(E_{R_n})}\Gamma^{J_i}_{bIl_i}(E_{R_n})
\end{equation}
and
\begin{equation}
\label{eq11} \Gamma^{J_i}_{\gamma
J_f}(E)=(\frac{E+\varepsilon_f}{E_{R_n}+\varepsilon_f})^{2L+1}\Gamma^{J_i}_{\gamma
J_f}(E_{R_n}),
\end{equation}
where $\Gamma^{J_i}_{bIl_i}(E_{R_n})$ and $\Gamma^{J_i}_{\gamma
J_f}(E_{R_n})$ are the experimental partial and radiative widths,
$\varepsilon_f$ is the proton binding energy of the bound state in
nucleus $A$, and L is the multipolarity of the gamma transition. The
penetrability $P_{l_i}(E)$ is expressed as
\begin{equation}
\label{eq12}
P_{l_i}(E)=\frac{kr_c}{F^2_{l_i}(k,r_c)+G^2_{l_i}(k,r_c)}.
\end{equation}

 The nonresonant amplitude can be calculated by
 \begin{eqnarray}
\label{eq13}
U^{NR}_{Il_iJ_fJ_i}&=&-(2)^{3/2}i^{l_i+L-l_f+1}e^{i(\omega_{l_i}-\phi_{l_i})}\frac{1}{\hbar
k}\mu_{Bb}^{L+1/2}
\nonumber\\
&&  \times \Big{[}\frac{Z_be}{m^L_b}+(-1)^L\frac{Z_Be}{m^L_B}\Big{]}\sqrt{\frac{(L+1)(2L+1)}{L}}\nonumber\\
&&  \times\frac{1}{(2L+1)!!}(k_\gamma
r_c)^{L+1/2}C_{J_fIl_f}F_{l_i}(k,r_c)\nonumber\\
&&\times G_{l_i}(k,r_c)W_{l_f}(2\kappa
r_c)\sqrt{P_{l_i}}(l_i0L0|l_f0)\nonumber\\
&&\times U(Ll_fJ_iI;l_iJ_f)J^\prime_L(l_il_f),
\end{eqnarray}
where
\begin{eqnarray}
\label{eq14} J^\prime_L(l_il_f)&=&
\frac{1}{r_c^{L+1}}\int^\infty_{r_c} dr \ r\frac{W_{l_f}(2\kappa
r)}{W_{l_f}(2\kappa
r_c)}\Big{[}\frac{F_{l_i}(k,r)}{F_{l_i}(k,r_c)}\nonumber\\
&&-\frac{G_{l_i}(k,r)}{G_{l_i}(k,r_c)}\Big{]}.
\end{eqnarray}
Here, $W_l(2\kappa r)$ is the Whittaker hypergeometric function,
$\kappa$ = $\sqrt{2\mu_{Bb}\varepsilon_f}$ and $l_f$ are the wave
number and relative orbital angular momentum of the bound state, and
$k_\gamma$ = $(E+\varepsilon_f)$/$\hbar c$ is the wave number of the
emitted photon.

The non-resonant amplitude contains the radial integral ranging only
from the channel radius $r_c$ to infinity since the internal
contribution is contained within the resonant part. Furthermore, the
R-matrix boundary condition at the channel radius $r_c$ implies that
the scattering of particles in the initial state is given by the
hard sphere phase. Hence, the problems related to the interior
contribution and the choice of incident channel optical parameters
do not occur. Therefore, the direct capture cross section only
depends on the ANC and the channel radius $r_c$.

\begin{figure}
\includegraphics[height =6.0 cm]{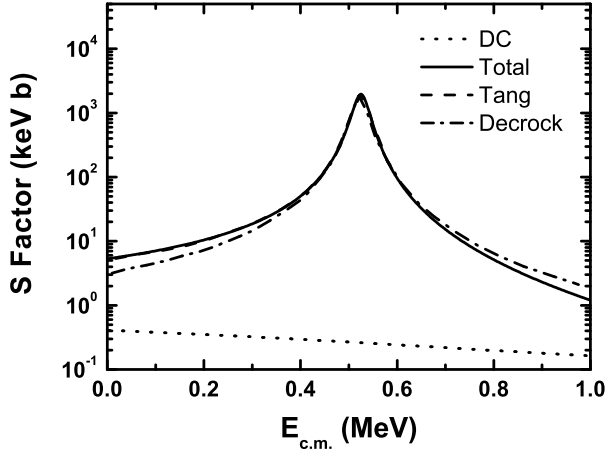}
\caption{\label{fig:sfactor}Astrophysical S-factors as a function of
$E_{\mathrm{c.m.}}$ for the $^{13}$N($p,\gamma$)$^{14}$O reaction.
The dotted line is the contributions from the direct proton capture.
The solid, dashed, and dashed-dotted lines indicate the total
S-factors from the present work, Refs. \cite{Tang04} and
\cite{Dec93}, respectively.}
\end{figure}

The astrophysical S-factor is related to the cross section by
\begin{equation}
\label{eq15}%
S(E)=E\sigma(E)\exp(E_{G}/E)^{1/2},
\end{equation}
where the Gamow energy $E_{G}=0.978Z^{2}_{1}Z^{2}_{2}\mu$ MeV, $\mu$
is the reduced mass of the system. According to the experimental ANC
(5.42 $\pm$ 0.48 fm$^{-1/2}$) from the present work, and the
resonance parameters ($E_{R}=527.9 \pm 1.7$ keV,
$\Gamma_{tot}(E_{R})=37.3 \pm 0.9$ keV, and
$\Gamma_{\gamma}(E_{R})=3.36 \pm 0.72$ eV) from Ref. \cite{Mag94},
the S-factors for direct and resonant captures can be then derived,
as demonstrated in Fig. \ref{fig:sfactor}.

\begin{table}
\caption{\label{tab3} The present total reaction rate for
$^{13}$N($p,\gamma$)$^{14}$O, $N_{A}<\sigma v>$
(cm$^{3}$mole$^{-1}$s$^{-1}$), as a function of temperature,
together with the previous results.}
\begin{ruledtabular}
{\renewcommand\baselinestretch{1.25}\selectfont
\begin{tabular}{cccc}
$T_{9}$ & Present work & Ref. [Tang] & NACRE \\
\hline 0.01 & $4.44\times10^{-22}$ &$4.18\times10^{-22}$&$2.01\times10^{-22}$\\
0.02 & $6.02\times10^{-16}$ &$5.72\times10^{-16}$&$2.78\times10^{-16}$\\
0.03 & $5.60\times10^{-13}$ &$5.35\times10^{-13}$&$2.63\times10^{-13}$\\
0.04 & $4.16\times10^{-11}$ &$3.98\times10^{-11}$&$1.99\times10^{-11}$\\
0.05 & $8.89\times10^{-10}$ &$8.53\times10^{-10}$&$4.34\times10^{-10}$\\
0.06 & $9.19\times10^{-9}$ &$8.84\times10^{-9}$&$4.58\times10^{-9}$\\
0.07 & $5.94\times10^{-8}$ &$5.72\times10^{-8}$&$3.02\times10^{-8}$\\
0.08 & $2.77\times10^{-7}$ &$2.67\times10^{-7}$&$1.44\times10^{-7}$\\
0.09 & $1.02\times10^{-6}$ &$9.86\times10^{-7}$&$5.43\times10^{-7}$\\
0.1 & $3.15\times10^{-6}$ &$3.04\times10^{-6}$&$1.71\times10^{-6}$\\
0.13 & $4.41\times10^{-5}$ &$4.27\times10^{-5}$&$2.56\times10^{-5}$\\
0.17 & $5.32\times10^{-4}$ &$5.16\times10^{-4}$&$3.34\times10^{-4}$\\
0.21 & $3.34\times10^{-3}$ &$3.24\times10^{-3}$&$2.22\times10^{-3}$\\
0.25 & $1.44\times10^{-2}$ &$1.39\times10^{-2}$&$9.85\times10^{-3}$\\
0.29 & $5.00\times10^{-2}$ &$4.84\times10^{-2}$&$3.47\times10^{-2}$\\
0.33 & $1.56\times10^{-1}$ &$1.51\times10^{-1}$&$1.11\times10^{-1}$\\
0.37 & $4.56\times10^{-1}$ &$4.41\times10^{-1}$&$3.41\times10^{-1}$\\
0.41 & $1.24\times10^{0}$ &$1.20\times10^{0}$&$9.91\times10^{-1}$\\
0.45 & $3.07\times10^{0}$ &$2.98\times10^{0}$&$2.60\times10^{0}$\\
0.49 & $6.87\times10^{0}$ &$6.69\times10^{0}$&$6.09\times10^{0}$\\
0.53 & $1.39\times10^{1}$ &$1.36\times10^{1}$&$1.28\times10^{1}$\\
0.57 & $2.59\times10^{1}$ &$2.54\times10^{1}$&$2.44\times10^{1}$\\
0.61 & $4.46\times10^{1}$ &$4.38\times10^{1}$&$4.27\times10^{1}$\\
0.65 & $7.20\times10^{1}$ &$7.09\times10^{1}$&$6.99\times10^{1}$\\
0.69 & $1.10\times10^{2}$ &$1.09\times10^{2}$&$1.08\times10^{2}$\\
0.73 & $1.60\times10^{2}$ &$1.58\times10^{2}$&$1.58\times10^{2}$\\
0.77 & $2.23\times10^{2}$ &$2.22\times10^{2}$&$2.22\times10^{2}$\\
0.81 & $3.01\times10^{2}$ &$2.99\times10^{2}$&$3.01\times10^{2}$\\
0.85 & $3.94\times10^{2}$ &$3.92\times10^{2}$&$3.95\times10^{2}$\\
0.89 & $5.02\times10^{2}$ &$5.00\times10^{2}$&$5.04\times10^{2}$\\
0.93 & $6.26\times10^{2}$ &$6.23\times10^{2}$&$6.30\times10^{2}$\\
0.97 & $7.64\times10^{2}$ &$7.59\times10^{2}$&$7.70\times10^{2}$\\
\end{tabular}
\par}
\end{ruledtabular}
\end{table}
Since the incoming angular momentum ($s$-wave) and the multipolarity
($E1$) of the direct and resonant capture $\gamma$-radiation are
identical, there is an interference between the direct and the
resonant captures. In this case, the total S-factor is calculated
with~\cite{Rol73}
\begin{equation}
\label{eq16} S_{tot}(E)=S_{dc}(E)+S_{res}(E) \pm
2[S_{dc}(E)S_{res}(E)]^{1/2}\cos(\delta),
\end{equation}
where $\delta$ is the resonance phase shift, which can be given by
\begin{equation}
\label{eq17} \delta=\arctan[{\Gamma_{p}(E) \over 2(E-E_{r})}].
\end{equation}
Generally, the sign of the interference in Eq. (\ref{eq16}) has to
be determined experimentally. However, it is possible sometimes to
infer this sign. The interference between the resonant and direct
capture contributions is constructive below the resonance energy and
destructive above it, which has been observed from the isospin
analog $^{13}$C($p,\gamma$)$^{14}$N$^{*}$ (2.31 MeV)
reaction~\cite{Dec93}. Recently, Tang et al. deduced constructive
interference below the resonance using an R-matrix method
\cite{Tang04}. Based on this interference pattern, the present total
S-factor is then obtained. Fig.~\ref{fig:sfactor} shows the
comparison of total S-factors from the present work, Refs.
\cite{Tang04} and \cite{Dec93}. Our updated total S-factors are
about 40\% higher than the previous ones in Ref. \cite{Dec93} at low
energies and is in good agreement with that in Ref.
\cite{Tang04}.\\
\section{The Astrophysical reaction rate}
The astrophysical reaction rate of $^{13}$N($p,\gamma$)$^{14}$O is
calculated with
\begin{eqnarray}
\label{eq18}%
N_A \langle\sigma v\rangle = N_A\big({8 \over \pi\mu}\big)^{1/2}{1
\over (kT)^{3/2}}\int^{\infty}_0 S(E) \nonumber\\
\times\exp\big[-({E_{G} \over E})^{1/2}-{E \over kT}\big]dE,
\end{eqnarray}
where $N_{A}$ is Avogadro constant. The updated rates are listed in
Table \ref{tab3}, together with the previous ones from
Ref.~\cite{Tang04} and NACRE's compilation. The results from the
three works agree with each other within a factor of 2 at low
temperature of $T$ $<$ 0.2 GK and are almost identical at higher
temperature of $T$ $>$ 0.7 GK.

\vspace{2mm}
\begin{figure}
\includegraphics[height =6.2 cm]{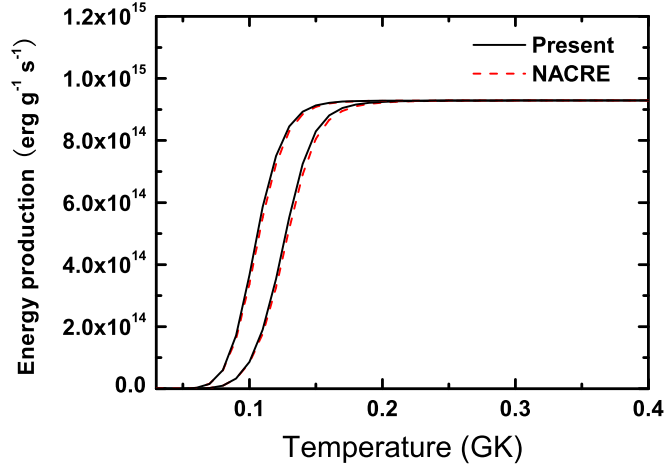}
\caption{\label{penergy}(Color online) Energy production rates of
the CNO and hot CNO cycles at $\rho$ = 500 (right) and 5000 (left)
$g$/cm$^3$ for novae with the update $^{13}$N($p,\gamma$)$^{14}$O
reaction rates from present work and NACRE's compilation.}
\end{figure}

The present total reaction rates as a function of temperature $T_9$
(in unit of $10^9$ K) are fitted with an expression used in the
astrophysical reaction rate library REACLIB \cite{Thi87},
\begin{widetext}
\begin{eqnarray}
\label{eq19}%
N_A \langle\sigma v\rangle &=& \exp[- 5.2635 + 0.0364
T_9^{-1}-21.5656 T_9^{-1/3} + 36.0575 T_9^{1/3} - 4.9432 T_9 +
0.3937 T_9^{5/3} \nonumber\\
&&- 9.7467 \ln(T_9)] + \exp[108.6965 + 0.6657 T_9^{-1} - 47.9051
T_9^{-1/3} - 59.4921 T_9^{1/3} + 5.0145 T_9  \nonumber\\
&&- 0.2488 T_9^{5/3} + 4.4288 \ln(T_9)].
\end{eqnarray}
\end{widetext}
The fitting errors are less than 5\% in the range from
$T_{9}=0.01$ to $T_{9}=10$.\\

For a given density $\rho$, the reaction network equations and the
energy source equation have the following forms:
\begin{equation}
\label{eq20}
\begin{array}{l}
\dot{Y_i}-F(Y_j,T)=0\\
\dot{\epsilon}+\sum\limits_iN_AM_ic^2\dot{Y_i}=0
\end{array}
\end{equation}
where $Y_i$ are the nuclear abundances, $\dot{\epsilon}$ is the
energy production rate per unit mass, $i,j$ = 1,2,$\cdots$,N, and N
is the number of nuclear species. $F$ denotes nonlinear functions of
the arguments, and $M_ic^2$ is the rest mass energy of species i in
MeV. At equilibrium, the abundances do not change with the time
approximately, i.e., $\dot{Y_i}$ $\simeq$ 0, the energy production
rate can then be calculated by substituting the reaction rates into
Eq.~(\ref{eq20}). Fig.~\ref{penergy} shows the energy productions of
CNO and hot CNO cycles at density $\rho$ = 500 and 5000 $g$/cm$^3$
for novae with the $^{13}$N($p,\gamma$)$^{14}$O reaction rates from
present work and NACRE's compilation. One can see that the hot CNO
cycle would begin to run earlier and produce more energy with our
updated $^{13}$N($p,\gamma$)$^{14}$O reaction rates. The present
result shows that about 5\% of additional energy could be produced
at the temperature range from 0.07 to 0.15 GK, which implies that
the evaluation of a novea may be affected.

\section{summary}
In summary, $^{13}$N($p,\gamma$)$^{14}$O is one of the key reactions
which trigger the onset of the hot CNO cycle. We have measured the
angular distribution of the $^{13}$N($d, n$)$^{14}$O reaction at
$E_\mathrm{c.m.}$ = 8.9 MeV, and deduced the nuclear ANC and
spectroscopic factor for the $^{14}$O ground state. The
astrophysical S-factors and reaction rates for $^{13}$N($p,
\gamma$)$^{14}$O are then extracted with the R-matrix approach. Our
result is in good agreement with that from the $^{14}$N
($^{13}$N,$^{14}$O)$^{13}$C transfer reaction by Tang et
al.~\cite{Tang04}. The reaction network calculations have been
performed with the updated $^{13}$N($p, \gamma$)$^{14}$O reaction
rates, the result shows that 5\% additional energy could be
generated through the CNO and hot CNO cycles at the typical
densities and temperature range from 0.07 to 0.15 GK for the novae,
this may affect the evaluation of novae.

\begin{acknowledgments}
This work was supported by the National Natural Science Foundation
of China under Grant Nos. 10375096, 10575136 and 10405035.
\end{acknowledgments}
\bibliography{n13dn}
\end{document}